\documentclass[superscriptaddress,showpacs,amssymb,10pt,reprint,aps,prd,longbibliography,nofootinbib,floatfix]{revtex4-2}

\usepackage{graphicx,epsfig,amssymb}
\usepackage{amsmath,amsfonts, times}
\usepackage{bm} 

\usepackage{epstopdf}
\usepackage[linktocpage,colorlinks]{hyperref}
\usepackage[caption=false]{subfig}
\usepackage[usenames]{color}     
\usepackage{natbib}
\usepackage{soul}
\usepackage[utf8x]{inputenc}
\usepackage[usenames]{color}

\definecolor{coolblack}{rgb}{0.0, 0.18, 0.39}
\definecolor{darkred}{rgb}{0.5,0,0}
\definecolor{darkgreen}{rgb}{0,0.5,0}
\definecolor{darkblue}{rgb}{0,0,0.5}
\definecolor{lapislazuli}{rgb}{0.15, 0.38, 0.61}
\definecolor{venetianred}{rgb}{0.78, 0.03, 0.08}
\definecolor{bleudefrance}{rgb}{0.19, 0.55, 0.91}
\definecolor{dogwoodrose}{rgb}{0.84, 0.09, 0.41}
\hypersetup{colorlinks=true, citecolor=darkgreen, linkcolor=darkblue, 
	urlcolor = blue}

\def\be{\begin{equation}}
\def\ee{\end{equation}}

\newcommand{\bea}{\begin{eqnarray}}
\newcommand{\eea}{\end{eqnarray}}
\newcommand{\ben}{\begin{enumerate}}
	\newcommand{\een}{\end{enumerate}}
\newcommand{\bi}{\begin{itemize}}
	\newcommand{\ei}{\end{itemize}}

\newcommand{\rt}{r_{\star}}

\def\ga{\mathrel{\raise.3ex\hbox{$>$\kern-.75em\lower1ex\hbox{$\sim$}}}}
\def\la{\mathrel{\raise.3ex\hbox{$<$\kern-.75em\lower1ex\hbox{$\sim$}}}}

\def\l{\left}
\def\r{\right}
\def\be{\begin{equation}}
\def\ee{\end{equation}}

\def\I_M{{I_{\scriptscriptstyle M\times M}}}

\def\be{\begin{equation}}
\def\ee{\end{equation}}
\def\bea{\begin{eqnarray}}
\def\eea{\end{eqnarray}}
\newcommand{\beq}{\begin{eqnarray}}
\newcommand{\eeq}{\end{eqnarray}}

\begin{document}

\title{Parametrized black holes: Scattering investigation}

\author{Renan B. Magalh\~aes}
	\email{rbmagalhaes22@hotmail.com}
	\affiliation{Programa de P\'os-Gradua\c{c}\~{a}o em F\'{\i}sica, Universidade 
		Federal do Par\'a, 66075-110, Bel\'em, Par\'a, Brazil.}
	
	\author{Luiz C. S. Leite}
	\email{luizcsleite@ufpa.br}
	\affiliation{Programa de P\'os-Gradua\c{c}\~{a}o em F\'{\i}sica, Universidade 
		Federal do Par\'a, 66075-110, Bel\'em, Par\'a, Brazil.}
	\affiliation{Campus Altamira, Instituto Federal do Par\'a, 68377-630, Altamira, Par\'a, Brazil.}
	
	\author{Lu\'is C. B. Crispino}
	\email{crispino@ufpa.br}
	\affiliation{Programa de P\'os-Gradua\c{c}\~{a}o em F\'{\i}sica, Universidade 
		Federal do Par\'a, 66075-110, Bel\'em, Par\'a, Brazil.}

\begin{abstract}
We study the scattering of light-like geodesics and massless scalar waves by a static Konoplya-Zhidenko black hole, considering the case that the parametrized black hole solution contains a single deformation parameter. By performing a geodesic analysis, we compute the classical differential scattering cross section and probe the influence of the deformation parameter on null trajectories. Moreover, we investigate the propagation of a massless scalar field in the vicinity of the static Konoplya-Zhidenko black hole and use the plane waves formalism to compute the differential scattering cross section. We confront our numerical results in the backward direction with the glory approximation, finding excellent agreement. We compare the results for the deformed black hole with the Schwarzschild case, finding that the additional parameter has an important role in the behavior of the scattering process for moderate-to-high scattering angles. 

\end{abstract}

\date{\today}
	
\maketitle
\section{Introduction}\label{sec:int}
Astrophysical observations lead us to believe that at the center of galaxies there are supermassive black holes~\cite{Eckart:2017}. The image of a compact object surrounded by hot plasma in the center of the M87 galaxy~\cite{eht} and the gravitational wave detections~\cite{ligo,ligo1,ligo2} vitalized the debate of the validity of General Relativity in the strong-field regime, providing new channels to obtain information about compact objects in the Universe, which allow us to compare standard General Relativity black hole solutions (for instance, Schwarzschild and Kerr black holes) with possible deviations~\cite{Mizuno:2018,JCCH:2021,Shaikh:2021}.

Instead of considering particular black hole solutions of alternative theories of gravity, one can use a parametrized approach~\cite{johannsen}. Several para\-metrized (or deformed) black holes were introduced and analyzed in the literature with different motivations~\cite{johannsen,cardoso:2014,Rezzola:2014,cardoso:2019,cardoso:2019ii}. In this work, we will consider the static Konoplya-Zhidenko solution~\cite{KZ}, which can be treated as a vacuum solution of some unknown alternative theory of gravity. The Konoplya-Zhidenko parametrization follows from changing the mass term of General Relativity solutions by adding a static deformation, such that the resulting metric naturally violates the no-hair theorems~\cite{costa}. 

The presence of the Konoplya-Zhidenko additional parameter changes sharply the spacetime in the strong-field regime~\cite{wang:2017}, such that physical processes are influenced and the corresponding outcomes may deviate from General Relativity predictions. These deviations from General Relativity are noticed both in the rotating and non-rotating versions of the Konoplya-Zhidenko spacetime. The shadow~\cite{wang:2017}, the strong gravitational lensing~\cite{wang:2016}, and the energy extraction process~\cite{long:2018} were studied for the rotating version of the solution. Additionally, the massless scalar field absorption was analyzed in the static case~\cite{MLC:2020PLB}.

As it has been pointed out in the literature~\cite{matzer:1968,fabbri,unruh,Futterman::book,CDHO:2011,MC:2014,CDHO:2015} scattering processes play a remarkable role in black hole physics. In order to fully consider field scattering, we need to solve a wave equation with suitable boundary conditions, and analyze how an effective potential influences the propagation of the incoming wave. It is possible to use the scattering analysis to extract information about black hole parameters, as their charge~\cite{CDHO2014,BODC:2014,BC:2016} and angular momentum~\cite{LDC:2017,LDC:2018,LBC:2019}. Therefore, one can perform a scattering investigation to obtain information about additional parameters beyond those allowed by the no-hair paradigm~\cite{MLC:2020EPJC,MLC:2021PRD}. 

Considering a spinless field in a Konoplya-Zhidenko deformed black hole background, we analyze the scattering process through the partial waves approach to probe how the additional parameter influences the scattering of massless scalar waves. We compare the numerical results of the differential scattering cross section obtained via the partial waves approach with the classical (geodesic) and semiclassical (glory) scattering cross sections. In particular, for large scattering angles ($\theta \approx \pi$), we find excellent agreement between the partial waves approach and the glory approximation.  

The remaining of this paper is organized as follows. In Sec.~\ref{sec:BHParametrization} we briefly present a parametrized black hole solution -- the static Konoplya-Zhidenko black hole spacetime. In Sec.~\ref{sec:scatt} we perform a scattering investigation of the static Konoplya-Zhidenko black hole. In particular, we discuss the null geodesic analysis, the glory approximation, and the partial waves approach. Furthermore, we present a selection of our numerical results and compare them with the analytical approximation. We conclude with our final remarks in Sec.~\ref{sec:remarks}. Throughout this paper we make use of natural units $G=c=\hslash=1$ and metric signature ($+$, $-$, $-$, $-$).

\section{BH parametrization}\label{sec:BHParametrization}
\begin{figure}[h]
\includegraphics[width=\columnwidth]{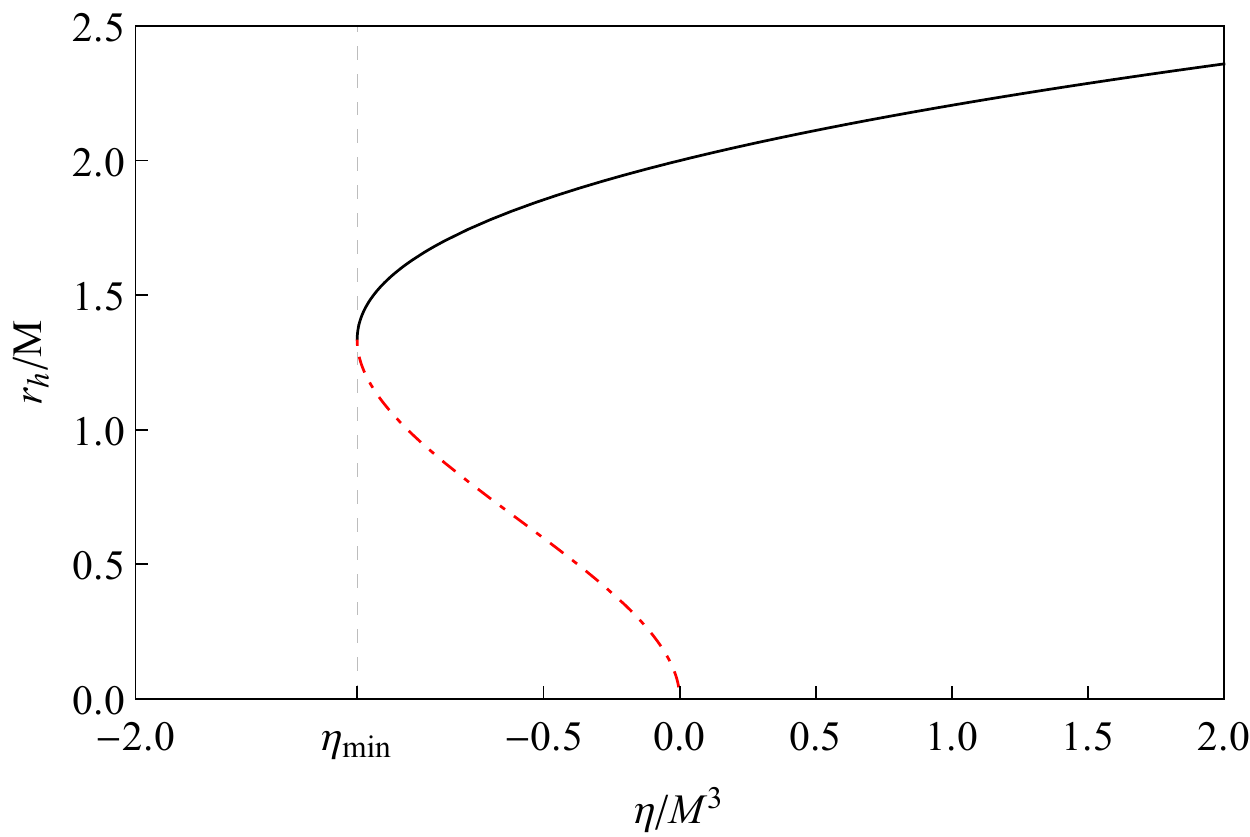}
\caption{The solid (black) curve gives the event horizon position as a function of the deformation parameter. The dot-dashed (red) curve represents a second (inner) horizon which appears for negative values of the deformation parameter. We notice that there is a minimum value of the deformation parameter, namely $\eta/M^3=\eta_{\text{min}}\equiv -32/27$, for the line element~\eqref{eq:parametrization} to describe a black hole.}
\label{fig:event_horizon_location}
\end{figure}

We consider the parametrization proposed in Ref.~\cite{KZ}, in which the black hole line element is modified by including additional parameters in the mass-term, the so-called deformation parameters. The line element of the non-spinning version of this parametrized black hole solution is given by~\cite{MLC:2020PLB}
\begin{equation}
\label{eq:KZ_lineelement}
ds^2 = f_{KZ}(r)dt^2-\dfrac{1}{f_{KZ}(r)}dr^2-r^2d\Omega^2,
\end{equation}
where $d\Omega^2\equiv d\theta^2+\sin^2\theta d\phi^2$ is the line element of the unit sphere and $f_{KZ}(r)\equiv 1-2M(r)/r$. In order to include additional parameters the mass-term $M(r)$ is chosen to be
\begin{equation}
\label{eq:mass_term}
M(r) = M+\dfrac{1}{2}\sum_{i=0}^{\infty}\dfrac{\eta_{i}}{r^i},
\end{equation}
where $\eta_i$ are the deformation parameters. 
If all $\eta_i$ vanish, 
the line element~\eqref{eq:KZ_lineelement} reduces to the Schwarzschild one. Within the family of parametrized black holes~\eqref{eq:KZ_lineelement}, one can chose a class of solutions with the same post-Newtonian asymptotics of Schwarzschild $(\beta=\gamma=1)$~\cite{KZ}. Throughout this paper we consider a parametrization that satisfies the latter constraint, namely: 
\begin{equation}
\label{eq:parametrization}
ds^2 = \left(1-\dfrac{2M}{r}-\dfrac{\eta}{r^3}\right)dt^2-\left(1-\dfrac{2M}{r}-\dfrac{\eta}{r^3}\right)^{-1}dr^2-r^2d\Omega^2,
\end{equation}
which can be obtained by assuming that $\eta_i = \eta \delta_{i\,2}$, in Eq.~\eqref{eq:mass_term}. This particular parametrization describes a black hole with two parameters, namely $M$ and $\eta$, and the relation between the event horizon position and the ADM mass $M$ is different from the Schwarzschild case~\cite{long:2018,MLC:2020PLB}. We show the event horizon position as a function of the deformation parameter in Fig.~\ref{fig:event_horizon_location}. We notice that the location of the event horizon increases monotonically, as we enhance the value of the deformation parameter, so that the area of the black hole increases for bigger values of the deformation parameter~\cite{MLC:2020PLB}.
\section{Scattering investigation}\label{sec:scatt}
In this section we investigate the (massless) scalar scattering by static Konoplya-Zhidenko parametrized black holes. To do so, we consider a massless spinless wave impinging from infinity. 
We present the key equations to analyze the scattering phenomena. We start by studying the high-frequency regime, making use of the geodesic and glory approximations, where some behaviors of the scalar wave scattering are foreshadowed. After that, by considering the partial waves approach, we study the scattering of massless scalar waves by parametrized black holes. 

\subsection{Null geodesic analysis}\label{sec:scatt_class}
\begin{figure}[h]
\includegraphics[width=\columnwidth]{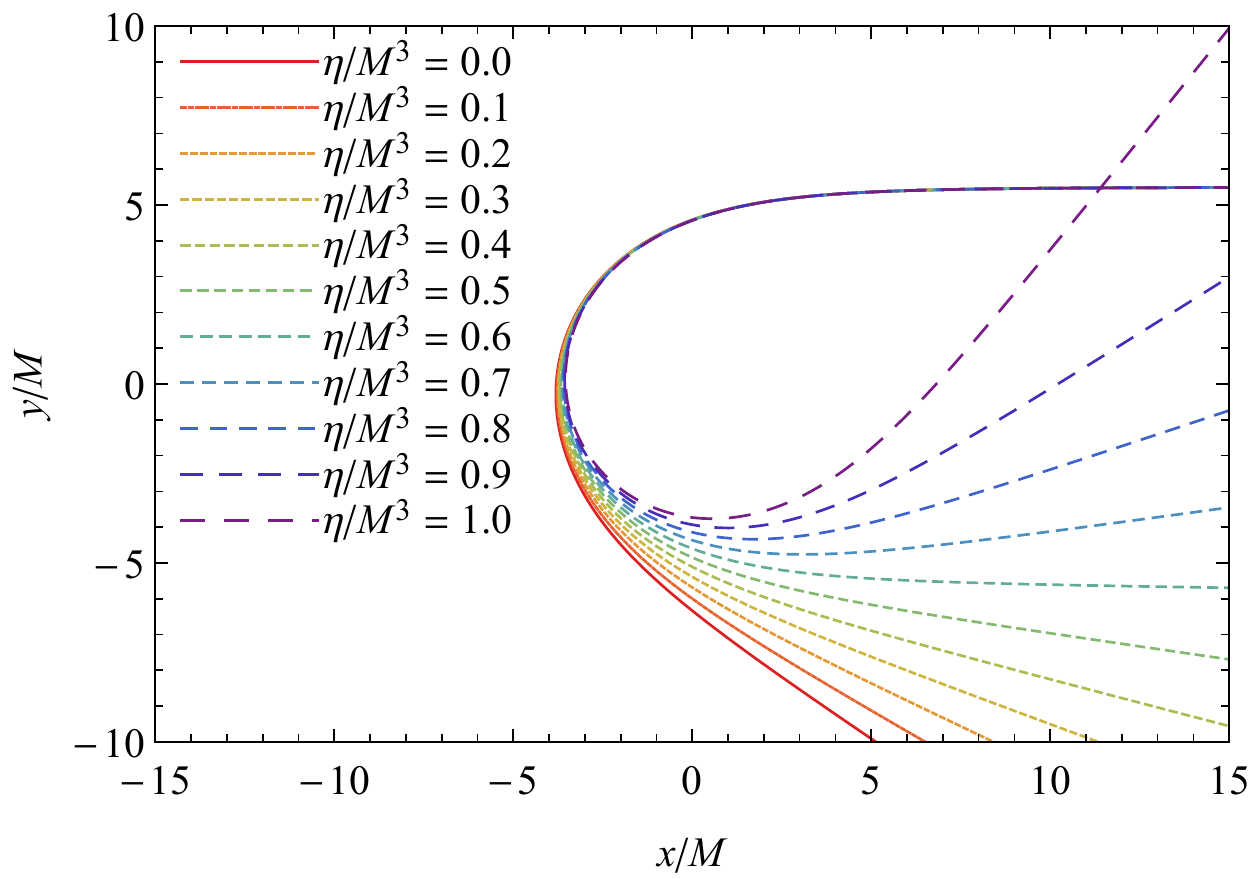}
\caption{Geodesics being bent in the parametrized black hole spacetime~\eqref{eq:parametrization}, for different values of the deformation parameter $\eta$. The impact parameter is chosen to be $b=5.5M$. We also show the Schwarzschild case ($\eta/M^3=0.0$).}
\label{fig:null_geodesics_KZ}
\end{figure}

In order to perform the geodesic analysis, we start by writing the orbit equation. 
Due to the spherical symmetry, we can, without loss of generality, consider the motion in the equatorial plane $(\theta=\pi/2)$, namely
\begin{equation}
\label{eq:orbit_equation}
\left(\dfrac{du}{d\phi}\right)^2 = \dfrac{1}{b^2}-u^2+2Mu^3+\eta u^5,
\end{equation}
where $u=1/r(\phi)$ and $b$ is the impact parameter, i.e, the perpendicular distance between the path of a particle (here null geodesics) and the scattering center (here black holes). One can differentiate Eq.~\eqref{eq:orbit_equation} and solve it with the appropriate boundary conditions to obtain the null geodesics in the parametrized black hole spacetime~\eqref{eq:parametrization}. In Fig.~\ref{fig:null_geodesics_KZ} we show a selection of null geodesics in different parametrized black hole spacetimes, and compare them with the Schwarzschild case. We notice that black holes with bigger deformation parameter have strong influence in the geodesic motion.  

\begin{figure}[h]
\includegraphics[width=\columnwidth]{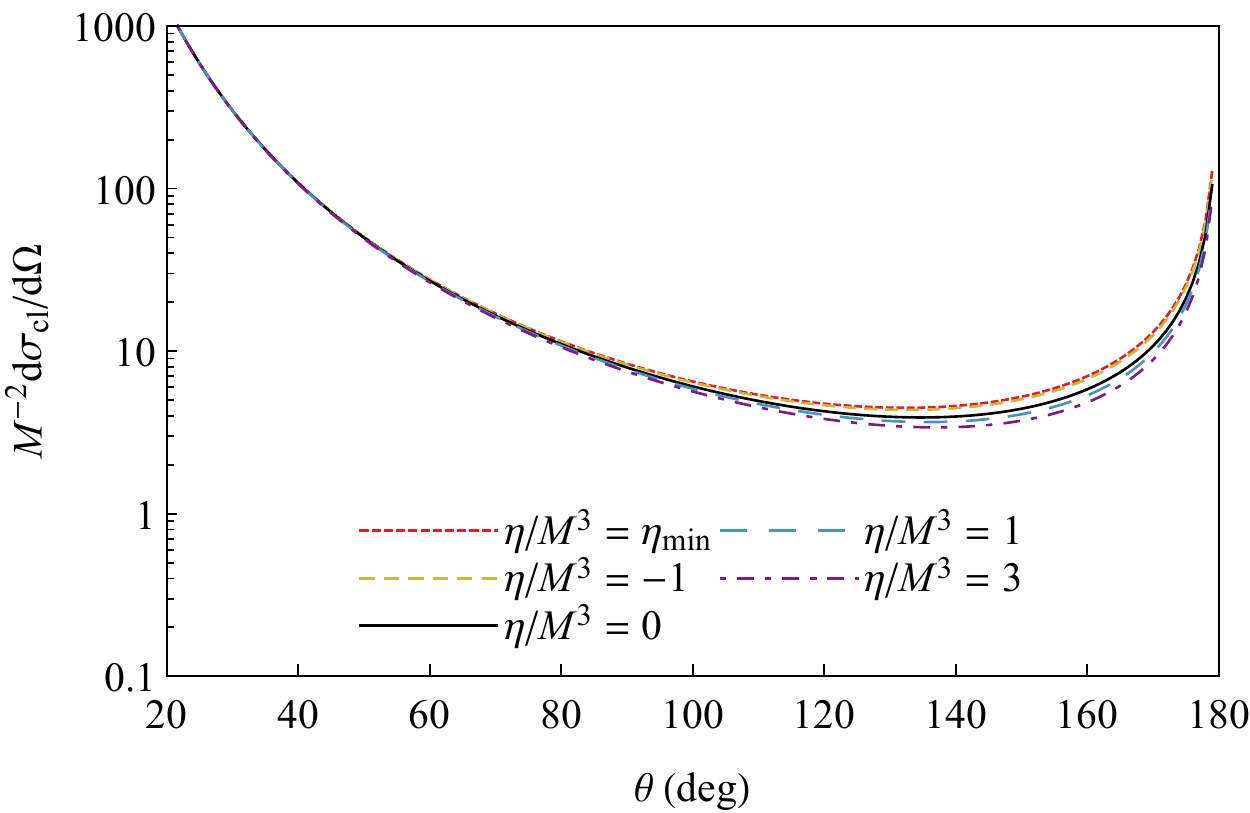}
\caption{Comparison between the classical scattering cross section of the parametrized black holes~\eqref{eq:parametrization}, for $\eta/M^3=\eta_{\text{min}}$, $-1$, 1 and 3, with the Schwarzschild case $(\eta/M^3=0)$.}
\label{fig:scatt_class_eta}
\end{figure}

The (classical) scattering cross section, giving the scattered flux in a specific direction, can be written as~\cite{collins:1973}
\begin{equation}
\label{eq:class_scatt}
\dfrac{d\sigma_{cl}}{d\Omega} = \csc \theta \sum_n b(\theta)\left\vert\dfrac{db(\theta)}{d\theta}\right\vert,
\end{equation}
where $b(\theta)$ is the impact parameter associated with a given scattering angle $\theta$. The sum in Eq.~\eqref{eq:class_scatt} takes into account the fact that particles can undergo $n$ $\left(\in \mathbb{N}\right)$ loops before being scattered through a deflection angle $\Theta$ (related to the scattering angle by $\theta=|\Theta-2n\pi|$). We shall see that Eq.~\eqref{eq:class_scatt} describes very well the massless scalar scattering cross section for small angles. In Fig.~\eqref{fig:scatt_class_eta} we exhibit the classical scattering cross section for some parametrized black holes~\eqref{eq:parametrization}, obtained by Eq.~\eqref{eq:class_scatt}. We note that black holes with smaller deformation parameters have a bigger classical scattering cross section for moderate-to-high scattering angles.
\subsection{Glory approximation}
Just as in Optics, black hole wave scattering gives rise to diffraction effects. The so-called glory is a bright spot or halo formed in the backward direction ($\theta=\pi$). The semi-classical approximation of the glory scattering cross section for scalar waves impinging upon spherically symmetric black holes is given by~\cite{matzner:1985} 
\begin{equation}
\label{eq:glory_app}
\dfrac{d\sigma_{\text{sc}}^{\text{g}}}{d\Omega}\Bigg\vert_{\theta\approx\pi} \approx 2\pi \omega b^2_{g}\left\vert\dfrac{db}{d\theta}\right\vert_{\theta\approx\pi} [J_{0}(\omega b_g \sin\theta)]^2,
\end{equation}
where $b(\pi)\equiv b_g$ is the glory impact parameter and  $J_0(x)$  is the Bessel function of the first kind, of order 0. In the high-frequency limit ($\omega M \gg 1$), the approximation~\eqref{eq:glory_app} gives excellent results for the scalar scattering in the backscattered direction $(\theta=\pi)$. We shall see that, even for $\omega M \sim 1$, the glory approximation provides very good agreement with the numerical results at $\theta\sim\pi$.

\subsection{Partial waves approach}\label{sec:scatt_pw}
The equation for the massless scalar field in a curved spacetime is
\begin{equation}
\label{eq:wave_eq}
\nabla^{\mu}\nabla_{\mu}\Psi=0.
\end{equation}
Due to the spherical symmetry, one can decompose the scalar field $\Psi$ as
\begin{equation}
\label{eq:psi_decomposition}
\Psi=\dfrac{\psi_{\omega l}(r)}{r}Y_{lm}(\theta,\phi)e^{-i\omega t},
\end{equation}
where $Y_{lm}(\theta,\phi)$ are the spherical harmonics. Substituting Eq.~\eqref{eq:psi_decomposition} into Eq.~\eqref{eq:wave_eq} we obtain the Schr\"{o}dinger-like equation
\begin{equation}
\label{eq:schro_eq}
\dfrac{d^2\psi_{\omega l}}{d\rt^2}+\left(\omega^2-V_l(r)\right)\psi_{\omega l}=0,
\end{equation}
where $\rt$ is the so-called tortoise coordinate, defined implicitly by $dr/d\rt=f_{KZ}(r)$, and

\begin{equation}
\label{eq:eff_potential}
V_l(r) = \dfrac{f_{KZ}(r)}{r}\dfrac{df_{KZ}}{dr}+f_{KZ}(r)\dfrac{l(l+1)}{r^2} 
\end{equation}
is the effective potential. At both asymptotic regions (event horizon, $\rt\rightarrow  -\infty$, and infinity, $\rt\rightarrow  \infty$) the effective potential goes to zero~\cite{MLC:2020PLB}, such that we look for solutions with the following boundary conditions
\be
\psi_{\omega l}(\rt)\sim\l\{
\begin{array}{ll}
	e^{-i \omega \rt}+{\cal R}_{\omega l }e^{i \omega \rt},&\rt\to+\infty, \\
	{\cal T}_{\omega l }e^{-i\omega \rt},& \rt\to-\infty,
\end{array}\r.\label{eq:inmodes}
\ee
that is, a composition of ingoing and outgoing waves at infinity and a purely incoming wave at the event horizon. In the latter equation, the coefficients ${\cal R}_{\omega l }$ and ${\cal T}_{\omega l }$ can be related to the reflection and transmission coefficients, respectively, and satisfy the flux conservation relation
\be
|{\cal R}_{\omega l }|^2+|{\cal T}_{\omega l }|^2=1.
\ee

The scalar differential scattering cross section for spherically symmetric black holes, can be written as
\begin{equation}
\label{eq:dif_scat}
\dfrac{d\sigma_{\text{sc}}}{d\Omega} = \left\vert\hat{g}(\theta)\right \vert^2,
\end{equation}
where
\begin{equation}
\label{eq:amp_sca}
\hat{g}(\theta) = \dfrac{1}{2i\omega} \sum_{l=0}^{\infty} (2l+1)\left[(-1)^{l+1}\mathcal{R}_{\omega l}-1\right]P_{l}(\cos\theta)
\end{equation}
is the scattering amplitude~\cite{CDO}. Due to the long-range nature of the fields propagating in the black hole spacetime, the scattering amplitude has a poor convergence~\cite{folacci:2019sw}, and we have to consider a high number of partial waves to numerically describe the divergence of $\hat{g}(\theta)$ in the forward direction $(\theta=0)$. In order to circumvent this difficulty, some procedures can be implemented, like the series reduction method~\cite{stratton:2020} or by considering a sum over the Regge Poles~\cite{folacci:2019sw,folacci:2019gw}. Here we use the series reduction method to obtain our numerical results.

\begin{figure}
\includegraphics[width=\columnwidth]{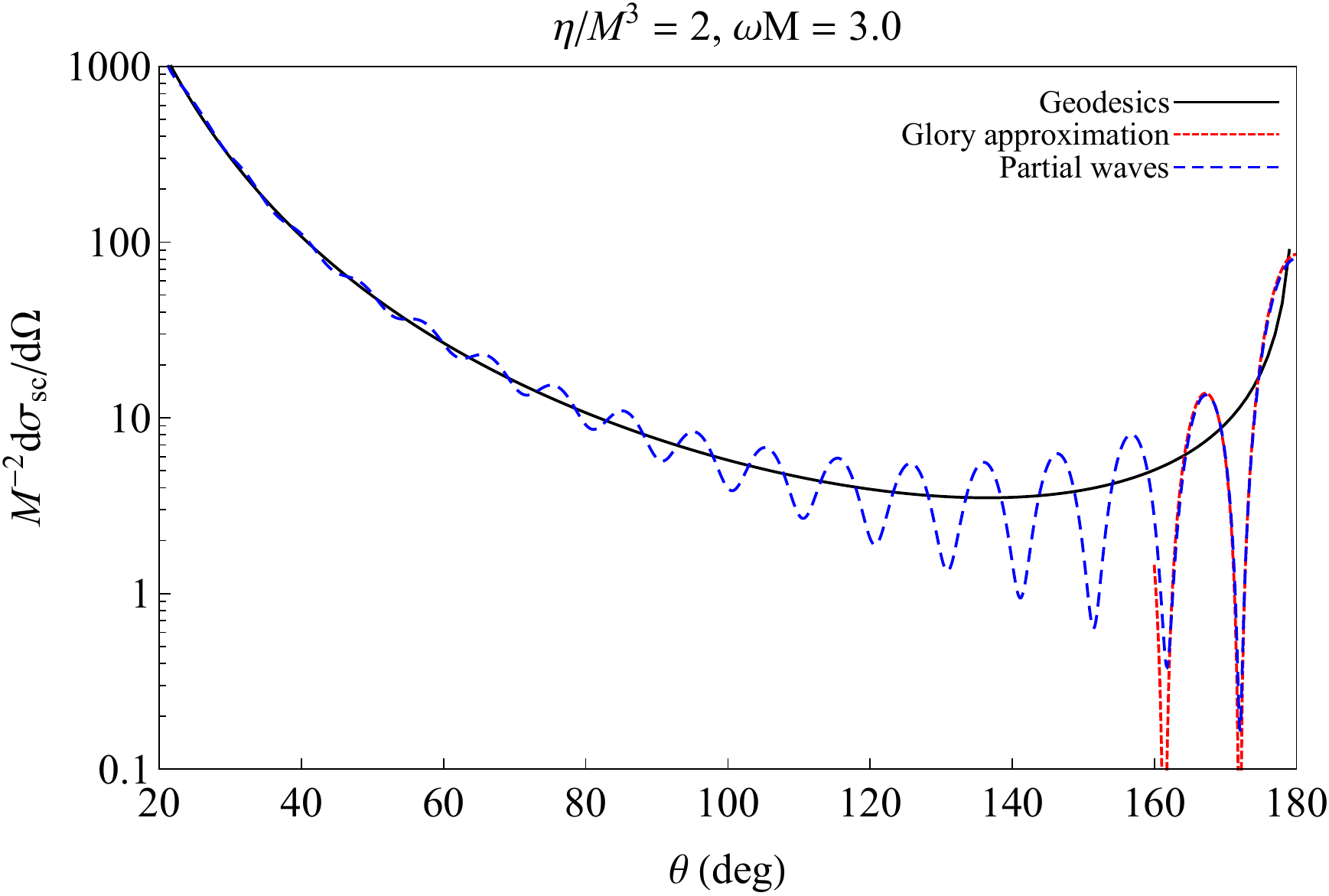}
\includegraphics[width=\columnwidth]{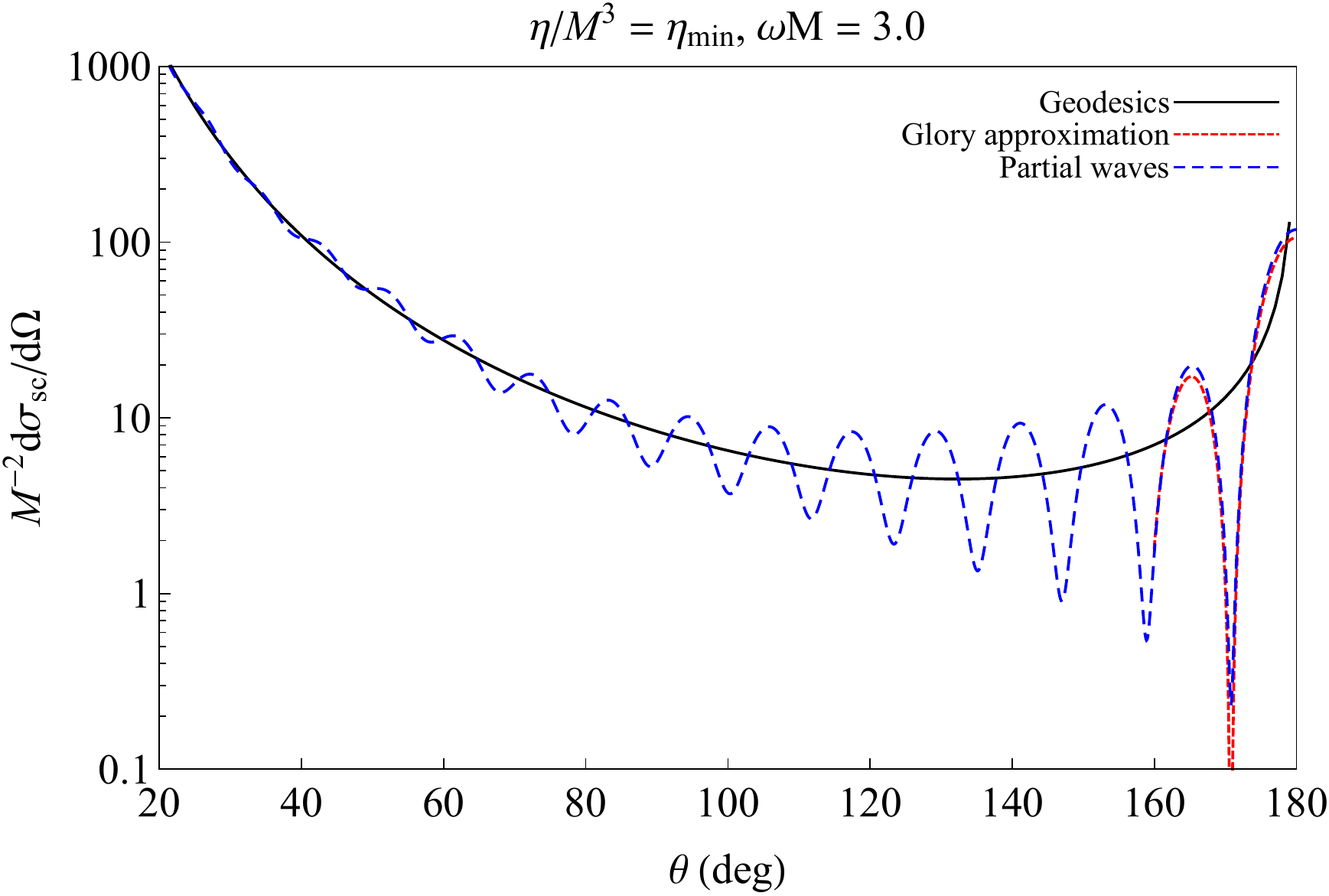}
\caption{Differential scattering cross section of parametrized black holes, obtained by geodesics analysis (black solid line), as well as by partial waves approach (blue dashed line). We also exhibit the glory scattering approximation (red dotted line). We consider $\eta/M^3=2$ (top panel), as well as $\eta/M^3=\eta_{\text{min}}$ (bottom panel), and $\omega M = 3$ for the wave analysis.}
\label{fig:fig:scatt_class_glory_pw}
\end{figure}

In Fig.~\ref{fig:fig:scatt_class_glory_pw} we plot the differential scattering cross section for a fixed value of frequency $(\omega M = 3)$ and two choices of the deformation parameter ($\eta/M^3 = 2$ and $\eta/M^3=\eta_{\text{min}}$). We compare the scattering cross section obtained via the partial wave approach (blue dashed line) with the ones obtained by the geodesic analysis (black solid line) and by the glory approximation (red dotted line). From Fig.~\ref{fig:fig:scatt_class_glory_pw}, we notice that in the forward direction $(\theta=0)$ the differential scattering cross section is well described by the geodesic scattering cross section. Additionally, for moderate-to-high values of the scattering angle, a remarkable interference pattern appears, and near the backward direction ($\theta \sim \pi$) the differential scattering cross section is in excellent accordance with the semiclassical glory approximation.

\begin{figure}[h]
\includegraphics[width=\columnwidth]{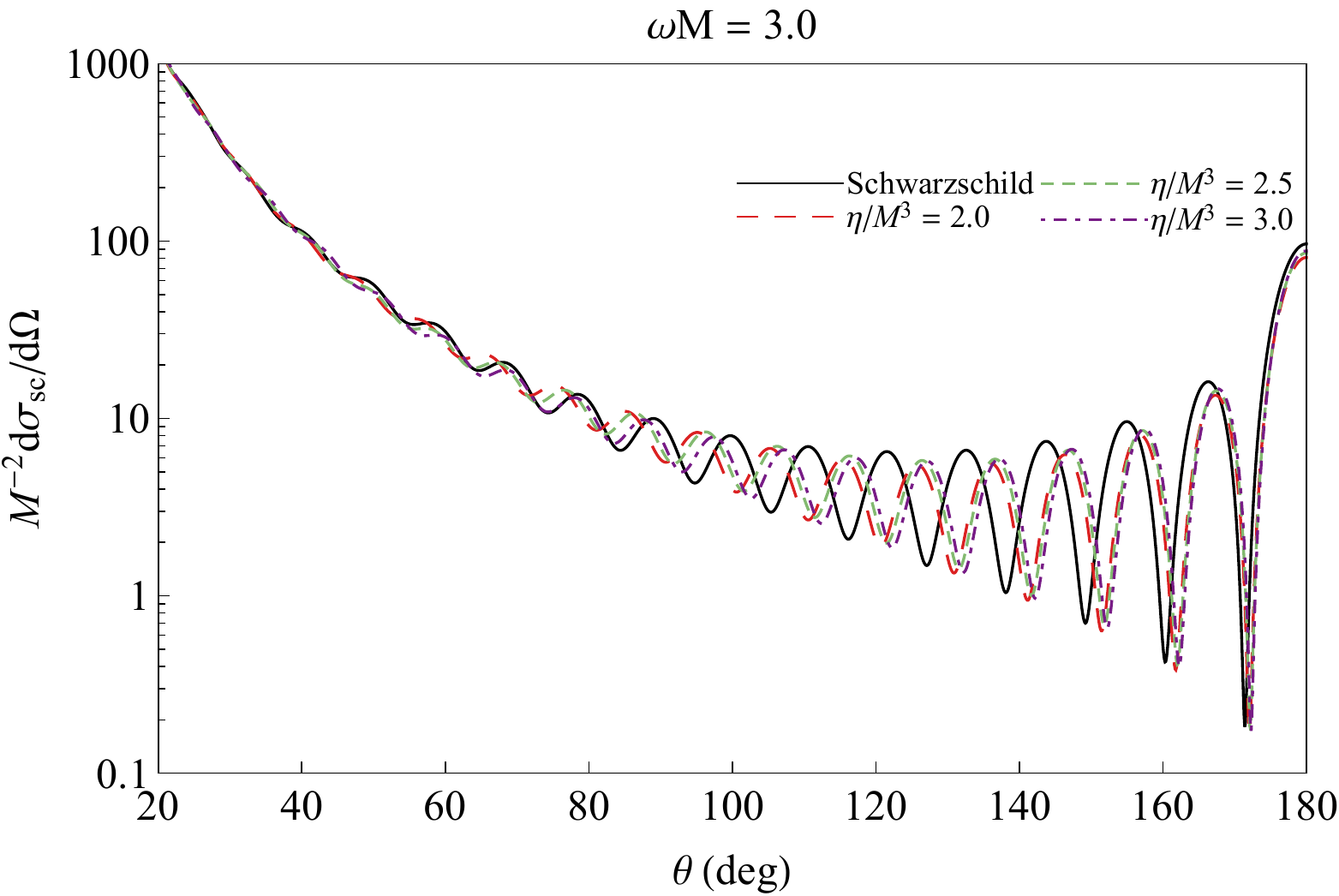}
\includegraphics[width=\columnwidth]{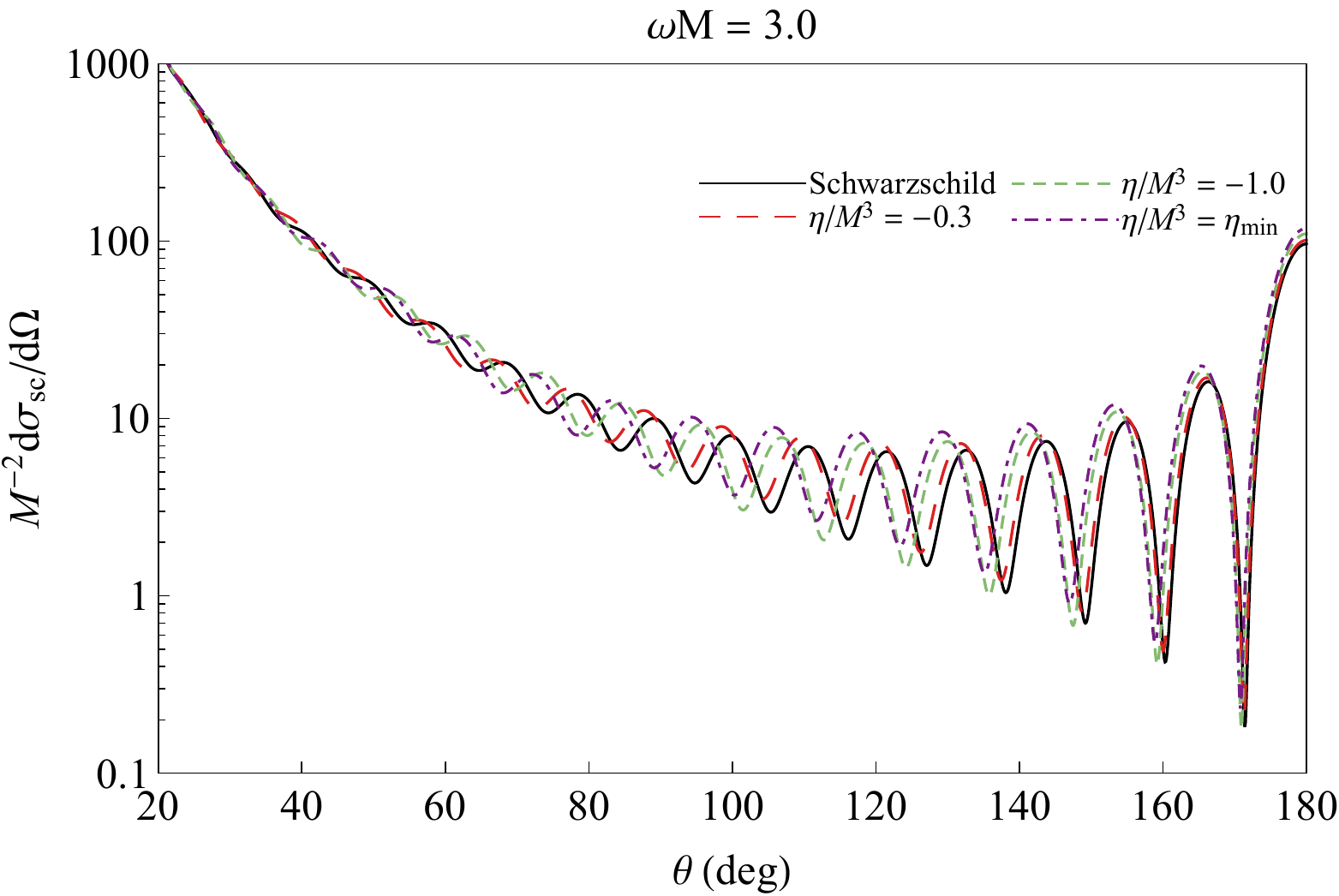}
\caption{Differential scattering cross section for some deformed black holes with positive (top panel) and negative (bottom panel) values of the deformation parameter. The Schwarzschild case is also shown, for comparison.}
\label{fig:scatt_w3_eta_range}
\end{figure}

\begin{figure}[h]
\includegraphics[width=\columnwidth]{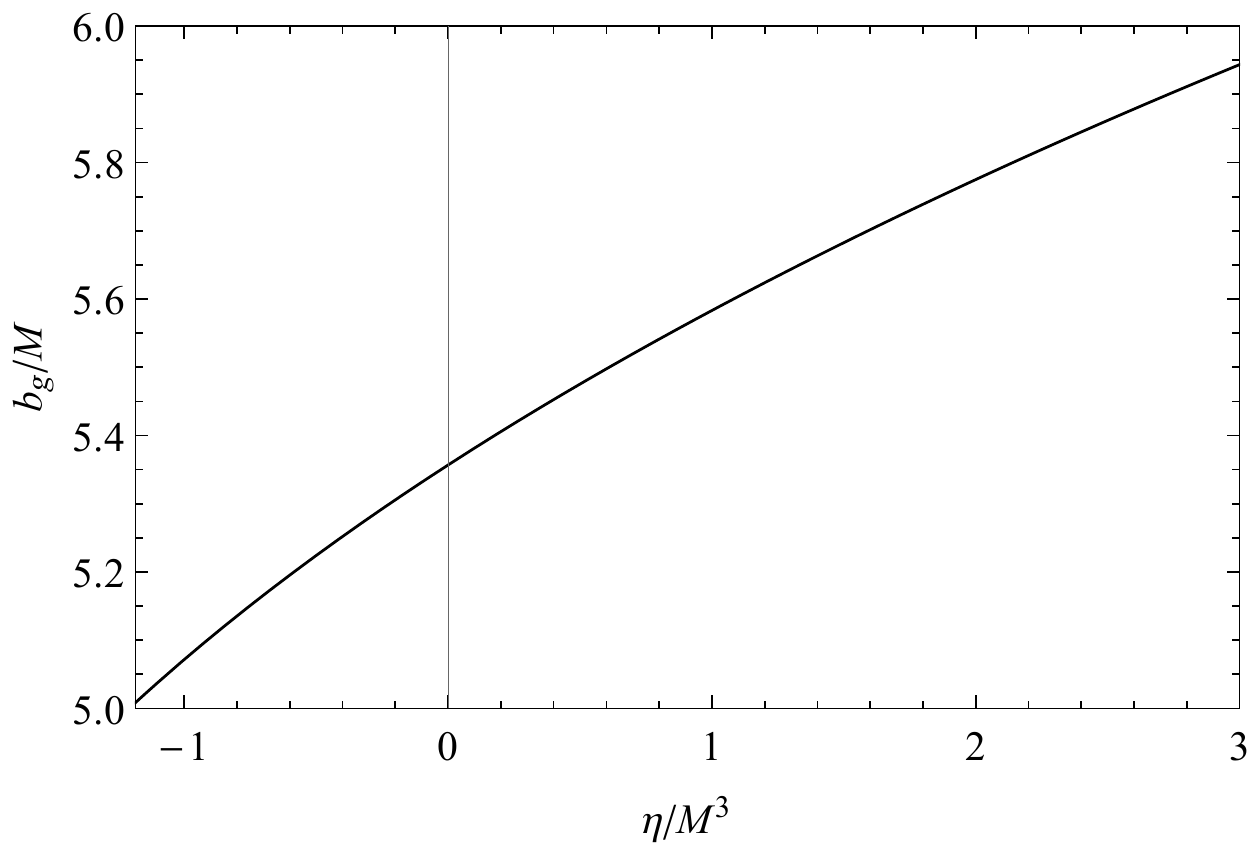}
\caption{Glory impact parameter as a function of the deformation parameter.}
\label{fig:bg_eta}
\end{figure}

In order to estimate the effects of the deformation parameter on the scattering of massless scalar waves, we plot in Fig.~\ref{fig:scatt_w3_eta_range} a selection of differential scattering cross sections for a fixed value of frequency ($\omega M = 3$) and several values of the deformation parameter (non-negative values on the top panel and non-positive values on the bottom panel). From Fig.~\ref{fig:scatt_w3_eta_range} we notice that by diminishing the value of the deformation parameter, the interference fringes width becomes larger. Additionally, the fringes are shifted to larger values of the scattering angle as we increase the value of the deformation parameter. The behavior of the fringes can be understood by analyzing the glory scattering~(namely, the scattering near the backward direction). We plot the behavior of the glory impact parameter in Fig.~\ref{fig:bg_eta},  and notice that $b_g$ increases monotonically as we enhance the value of the deformation parameter. By considering Eq.~\eqref{eq:glory_app}, we notice that the interference fringe widths are proportional to $1/(b_g\omega)$, so that by increasing the value of the deformation parameter or the value of the frequency, the fringe widths get narrower. 

\begin{figure}[h]
\includegraphics[width=\columnwidth]{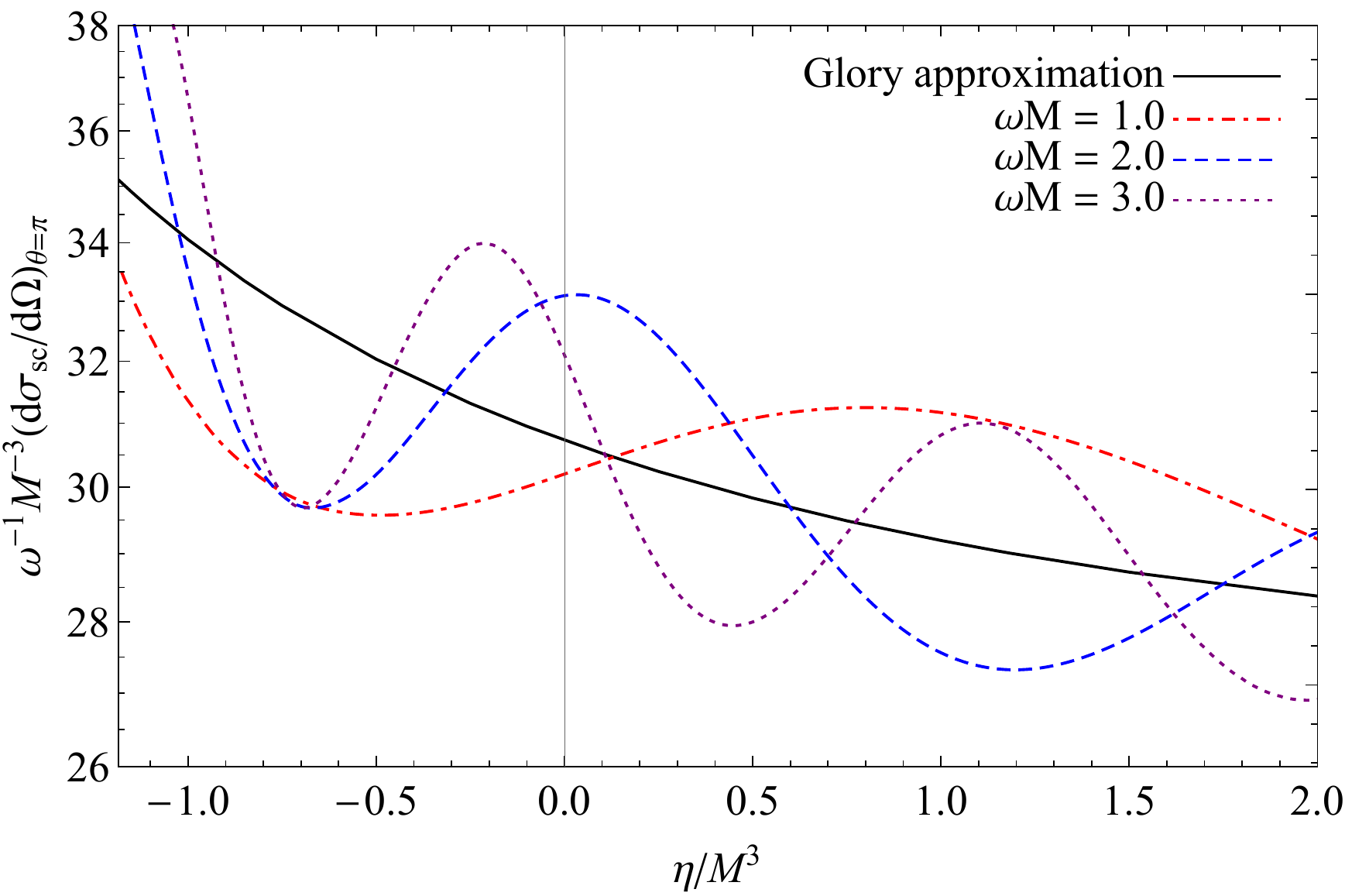}
\caption{Amplitude of the glory $(\theta=\pi)$ as a function of the deformation parameter $\eta$. We notice that the numerical results oscillate around the glory approximation (monotonically decreasing solid black curve).}
\label{fig:glory_amp}
\end{figure}

We can see from Fig.~\ref{fig:scatt_w3_eta_range} that the magnitude of the glory peak in the backward direction depends on the value of the deformation parameter. From Fig.~\ref{fig:fig:scatt_class_glory_pw} we note that the numerical value of the glory peak does not, in general, coincide with the peak of the glory approximation. We illustrate this difference in Fig.~\ref{fig:glory_amp}, from where we see that the numerical values of the glory peak oscillate around the semiclassical approximation (black solid line).

\section{Final remarks}\label{sec:remarks}
We have investigated the scattering of null geodesics
and massless scalar waves by a static Konoplya-Zhidenko
black hole containing a single extra parameter, and compared the results obtained with the well-known Schwarzschild case.

By analyzing how the additional parameter affects the geodesic motion, we have found that small deformation parameters have a weak influence on the null-like trajectories around black holes. 
With the increase of the deformation parameter, the critical impact parameter of the Konoplya-Zhidenko black holes also increases, and its shadow becomes bigger, reinforcing the results of Ref.~\cite{wang:2017}. We have computed the classical differential scattering cross section and noticed that the increase of the deformation parameter diminishes the differential scattering cross section for moderate-to-high scattering angles. 

We have studied how the additional parameter influences the scattering of massless scalar waves. In the forward direction ($\theta=0$), the differential scattering cross section computed by the partial waves approach degenerates into the classical scattering cross section. For small scattering angles, the role of the extra parameter is less relevant. However, for moderate-to-high scattering angles, the effect of the deformation parameter on the scattering process becomes more important. The semiclassical analysis allows us to understand the oscillatory pattern in the differential scattering cross section (see, for instance, Fig.~\ref{fig:fig:scatt_class_glory_pw}) as being due to the interference between rays orbiting the black hole in opposite senses. The semiclassical analysis also explains the role of the deformation parameter near the backward direction ($\theta=\pi$). As the deformation parameter increases, the glory impact parameter grows and the interference fringe widths get narrower. In the backward direction, we notice that there is an intensity peak (glory peak), which depends on the value of the deformation parameter.

The results reported here
reinforce
that the additional parameters in the static parametrized black hole solutions have a strong influence on physical processes near the event horizon. It should be interesting to extend the analysis presented here to rotating versions of deformed black holes in order to study how the extra parameters influence the physics near black holes with non-zero angular momentum.

\begin{acknowledgements}
The authors would like to acknowledge 
Conselho Nacional de Desenvolvimento Cient\'ifico e Tecnol\'ogico (CNPq)
and Coordena\c{c}\~ao de Aperfei\c{c}oamento de Pessoal de N\'ivel Superior (CAPES) -- Finance Code 001, from Brazil, for partial financial support. This research has also received funding from the European Union's Horizon 2020 research and innovation programme under the H2020-MSCA-RISE-2017 Grant No. FunFiCO-777740.
\end{acknowledgements}

{}
\end{document}